\documentclass[pra, aps, twocolumn, floatfix, showpacs]{revtex4}
\usepackage{graphicx, amsmath, amssymb, times}

\begin{document}
\title{Counterflow of spontaneous mass currents in trapped spin-orbit coupled Fermi gases}
\author{E. Doko$^1$, A. L. Suba{\c s}{\i}$^2$, and M. Iskin$^1$}
\affiliation{
$^1$Department of Physics, Ko\c c University, Rumelifeneri Yolu, 34450 Sar{\i}yer, Istanbul, Turkey. \\
$^2$Department of Physics, Faculty of Science and Letters, Istanbul Technical University, 34469 Maslak, Istanbul, Turkey.
}
\date{\today}

\begin{abstract}
We use the Bogoliubov-de Gennes formalism and study the ground-state phases of trapped 
spin-orbit coupled Fermi gases in two dimensions. Our main finding is that the presence
of a symmetric (Rashba-type) spin-orbit coupling spontaneously induces counterflowing 
mass currents in the vicinity of the trap edge, i.e. $\uparrow$ and $\downarrow$ 
particles circulate in opposite directions with equal speed. These currents flow even in
noninteracting systems, but their strength decreases toward the molecular BEC limit, 
which can be achieved either by increasing the spin-orbit coupling or the interaction strength. 
These currents are also quite robust against the effects of asymmetric spin-orbit couplings 
in $x$ and $y$ directions, gradually reducing to zero as the spin-orbit coupling becomes 
one dimensional. We compare our results with those of chiral $p$-wave superfluids/superconductors.
\end{abstract}

\maketitle

\section{Introduction}
\label{sec:intro}

The controversy over the expectation value of the intrinsic ground-state angular momentum
of the $A$-phase of superfluid $^3$He in a given container has a long history, and it has not
yet attained a universally agreed resolution~\cite{leggett}. For $N$ weakly-interacting particles, 
theoretical predictions for the BCS ground state vary orders of magnitude from 
$N |\Delta| \hbar / (2\varepsilon_F)$ to $N \hbar/2$, where $|\Delta|$ is the amplitude 
of the order parameter, i.e. pairing energy, and $\varepsilon_F$ is the Fermi energy. 
At a first glance, the former expectation, which is based on the observation that pairing 
affects not all but only a small fraction $(|\Delta|/\varepsilon_F) N$ of fermions, 
seems more intuitive. However, this subject has recently regained interest 
in condensed-matter and atomic and molecular physics communities due to its 
relevance to the possible $p_x + ip_y$-superconducting phase of Sr$_2$RuO$_4$~\cite{mackenzie} 
and atomic chiral $p$-wave superfluids~\cite{jin}, respectively, 
supporting for the latter expectation~\cite{stone, mizushima}. On one hand, 
these recent results seem intuitive at least in the strong fermion attraction 
limit of tightly-bound $p$-wave Cooper molecules, where all of the molecules 
can undergo BEC at zero temperature with each molecule having a microscopic 
angular momentum $\hbar$.
On the other hand, they seem counterintuitive in the weak fermion 
attraction limit of loosely bound and largely overlapping $p$-wave Cooper pairs, since 
these results imply that the angular momentum in the BCS ground state is 
macroscopically different from its vanishing value in the normal ground state, 
when the pairing energy becomes arbitrarily small. 
While this controversy still awaits for an experimental resolution, here we propose 
an alternative atomic system where the microscopic angular momentum of Cooper 
pairs can also give rise to a macroscopic one. With their unique advantages 
over the condensed-matter systems, it is plausible that the macroscopic 
angular momentum of this alternative system may be observed for the first time 
with cold atoms, and that this would also shed some light on the $^3$He controversy 
for which the basic mechanism is found to be similar. 

Cold atom systems have already proved to be versatile in simulating various 
many-body problems. For instance, one of the major achievements with atomic 
Fermi gases in the past decade is the unprecedented realization of the BCS-BEC 
crossover~\cite{review}, where the ground state of the system can be 
continuously tuned from the BCS to the BEC limit as a function of increasing 
fermion attraction with the turn of a knob. The main difference between the 
BCS-BEC crossover and the BCS theory is that the Cooper pairing is
allowed not only for fermions with energies close to the Fermi energy but also 
for all pairs with appropriate momenta~\cite{leggett, review}. 
Having established the basic tools, it is arguable that 
one of the most promising research directions to pursue with cold Fermi gases 
is the artificial spin-orbit gauge fields. Such a field has already been realized
with cold Bose gases~\cite{nist1, nist2} using a novel technique based on 
spatially-dependent optical coupling between internal states of the atoms, 
and the same technique is equally applicable for neutral fermionic atoms.
In fact, we have recently learned that the first generation of quantum 
degenerate spin-orbit coupled Fermi gases has recently been created 
with $^{40}$K atoms~\cite{zhainote}.

This immediate possibility of creating a spin-orbit coupled Fermi gas has already 
received tremendous theoretical interest in condensed-matter and atomic 
physics communities, exploring a number of exotic superfluid phases with 
balanced or imbalanced populations, at zero or finite temperatures, in two or three 
dimensions, etc.~\cite{gong, shenoy, zhai, hui, subasi, wyi, 
carlos, he, ghosh, yang, zhou, duan, zhang, liao}.
Motivated by these recent advances, here we study ground-state and finite-temperature 
phases of trapped spin-orbit coupled Fermi gases, and our main results are as follows. 
We find that the presence of a spin-orbit coupling (SOC) spontaneously induces 
counterflowing $\uparrow$ and $\downarrow$ mass currents in the vicinity 
of the trap edge~\cite{iskin-vortex}. We show that these currents flow even in noninteracting 
systems, and that they are quite robust against the effects of asymmetric SOC 
in $x$ and $y$ directions, gradually reducing to zero as the SOC becomes 
one dimensional. However, their strength decreases toward the molecular 
BEC limit, which can be achieved either by increasing the SOC or the 
interaction strength. We argue that the origin of spontaneously induced 
counterflowing mass currents in spin-orbit coupled Fermi gases can be 
understood via a direct correspondence with the chiral $p$-wave 
superfluids/superconductors, however, with some fundamental 
differences in between.

The rest of the paper is organized as follows. In Sec.~\ref{sec:background}, 
first we introduce the mean-field Hamiltonian that is used to describe the 
spin-orbit coupled Fermi gases on optical lattices, and then use the 
Bogoliubov-de Gennes (BdG) formalism to obtain the generalized 
self-consistency equations for the number of fermions, superfluid order 
parameter, and strength of circulating mass currents. The numerical solutions 
of this formalism are presented in Sec.~\ref{sec:numerics}, which is followed by a 
brief discussion on their experimental realization in Sec.~\ref{sec:disc}. 
We summarize our main findings in~\ref{sec:conc}.

\section{Theoretical Background}
\label{sec:background}

In this paper, we consider a harmonically trapped two-dimensional optical 
lattice in the presence of a non-Abelian gauge potential. In addition to the 
usual spin-conserving hopping terms, the main effect of such a gauge 
potential is the appearance of an additional spin-flipping hopping term in 
the single-particle kinetic energy term as discussed next.

\subsection{Hamiltonian}
\label{sec:ham}

Spin-orbit coupled Fermi gases on optical lattices can be described by 
the grand-canonical mean-field Hamiltonian,
\begin{align}
H = &-t \sum_{i, \hat{e}} \left(C_{i+\hat{e}}^\dagger \phi_{i+\hat{e},i} C_i + H.c.\right) \nonumber \\
& - \sum_i \left(\mu_\sigma + V_{\sigma i}^H - V_i^T\right) c_{\sigma i}^\dagger c_{\sigma i} \nonumber \\
&+ \sum_i \left(\Delta_i c_{\uparrow i}^\dagger c_{\downarrow i}^\dagger + \Delta_i^* c_{\downarrow i} c_{\uparrow i}\right),
\end{align}
where the operator $c_{\sigma i}^\dagger$ ($c_{\sigma i}$) creates (annihilates) a
pseudo-spin $\sigma = \{ \uparrow, \downarrow\}$ fermion at lattice site $i$, the spinor 
$C_i^\dagger = (c_{\uparrow i}^\dagger, c_{\downarrow i}^\dagger)$ denotes the 
fermion operators collectively, $\hat{e} = \{\hat{x}, \hat{y}\}$ allows only nearest-neighbor 
hopping, and $H.c.$ is the Hermitian conjugate.
For a generic gauge field $\mathbf{A}=(\alpha \sigma_y, -\beta\sigma_x)$, 
where $\sigma_e$ is the Pauli matrix and $\{\alpha, \beta\} \ge 0$ are independent 
parameters characterizing both the strength and the symmetry of the spin-orbit coupling, 
the $\uparrow$ and $\downarrow$ fermions gain
$
\phi_{i+\hat{x}, i} = e^{-i\alpha \sigma_y}
$
phase factors for hopping in the positive $x$ direction and
$
\phi_{i+\hat{y}, i} = e^{i\beta \sigma_x}
$
phase factors for hopping in the positive $y$ direction.
Note that the spin-conserving and spin-flipping hopping terms are proportional, 
respectively, to $\cos \alpha$ and $\sin \alpha$ in the $x$ direction, and to $\cos \beta$ 
and $\sin \beta$ in the $y$ direction. The off-diagonal coupling 
$\Delta_i$ is the mean-field superfluid order parameter to be specified below, and
$V_{\sigma i}^H = g n_{-\sigma i}$ is the Hartree term where $g \ge 0$ is the strength 
of the onsite attractive interaction and $n_{\sigma i}$ is the filling of $\sigma$ fermions
at site $i$. Here, $(-\uparrow) = \downarrow$ and vice versa.
We introduce $\sigma$-dependent chemical potentials $\mu_\sigma$ to fix the 
number of $\sigma$ fermions independently, but we assume both $\uparrow$ 
and $\downarrow$ fermions feel the same trapping potential $V_i^T = V_0 r_i^2/2$, 
where the distance $r_i$ is measured from the center in our $L \times L$ lattice. 
Next, we solve this Hamiltonian via BdG formalism.

\subsection{Bogoliubov-de Gennes formalism}
\label{sec:bdg}

The BdG equations are obtained by diagonalizing the quadratic Hamiltonian 
given above via a generalized Bogoliubov-Valatin transformation, 
leading to a $4L^2 \times 4L^2$ matrix-eigenvalue problem,
\begin{align}
\label{eqn:bdg.matrix}
\sum_{j} \left( \begin{array}{cccc}
T_{\uparrow \uparrow} & T_{\uparrow \downarrow} & 0 & \Delta \\
T_{\downarrow \uparrow} & T_{\downarrow \downarrow} & -\Delta & 0 \\ 
0 & -\Delta^* & -T_{\uparrow \uparrow}^* & -T_{\uparrow \downarrow}^* \\ 
\Delta^*& 0 & -T_{\downarrow \uparrow}^* & -T_{\downarrow \downarrow}^* 
\end{array} \right)_{ij}
&
\left( \begin{array}{c}
u_{nj}^\uparrow \\
u_{nj}^\downarrow \\
v_{nj}^\uparrow \\
v_{nj}^\downarrow
\end{array} \right) 
= \varepsilon_n
\left( \begin{array}{c}
u_{ni}^\uparrow \\
u_{ni}^\downarrow \\
v_{ni}^\uparrow \\
v_{ni}^\downarrow
\end{array} \right),
\end{align}
where $u_{ni}^\sigma$ and $v_{ni}^\sigma$ are the components of the $n$th quasiparticle
wave function at site $i$, and $\varepsilon_n \ge 0$ is the corresponding energy eigenvalue.
While the off-diagonal couplings are $\Delta_{ij} = \Delta_i \delta_{ij}$ diagonal in site indices 
since we consider onsite interactions only, the nearest-neighbor hopping and onsite 
energy terms can be written compactly as 
$
T_{\sigma\sigma'}^{ij} = -t_{\sigma \sigma'}^{ij} - (\mu_\sigma + V_{\sigma i}^H - V_i^T) \delta_{ij} \delta_{\sigma \sigma'},
$
where $\delta_{ij}$ is the Kronecker delta. Here, the non-vanishing nearest-neighbor hopping 
elements are 
$t_{\sigma\sigma}^{i,i+\hat{x}} = t \cos \alpha$ and 
$t_{\uparrow\downarrow}^{i,i+\hat{x}} = - t_{\downarrow\uparrow}^{i,i+\hat{x}} = -t \sin \alpha$
for the positive $x$ direction, 
and $t_{\sigma\sigma}^{i,i+\hat{y}} = t \cos \beta$ and 
$t_{\uparrow\downarrow}^{i,i+\hat{y}} = t_{\downarrow\uparrow}^{i,i+\hat{y}} = i t \sin \beta$
for the positive $y$ direction.
Note that the hopping in the negative $x$ and $y$ directions are simply the 
Hermitian conjugates. 

Equation~(\ref{eqn:bdg.matrix}) needs to be solved simultaneously with the order parameter
$
\Delta_i = g \langle c_{\uparrow i} c_{\downarrow i} \rangle
$
and number 
$
n_{\sigma i} = \langle c_{\sigma i}^\dagger c_{\sigma i} \rangle
$
equations for a self-consistent set of $\Delta_i$ and $\mu_\sigma$ solutions. 
Here, $\langle \cdots \rangle$ is a thermal average. 
Using the Bogoliubov-Valatin transformation, we obtain
\begin{align}
\label{eqn:op}
\Delta_i &= g \sum_n \Big[ u_{n i}^\uparrow (v_{n i}^ \downarrow)^* f(-\varepsilon_n)
+ u_{n i}^ \downarrow (v_{n i}^\uparrow)^* f(\varepsilon_n) \Big], \\
\label{eqn:number}
n_{\sigma i} &= \sum_n \Big[ |u_{n i}^\sigma|^2 f(\varepsilon_n) + |v_{n i}^\sigma|^2 f(-\varepsilon_n) \Big],
\end{align}
where $f(x)=1/(e^{x/T}+1)$ is the Fermi-Dirac distribution function and $T$ is the
temperature. Here, we set the Boltzmann constant $k_B$ to unity.
Note that 
$
0 \le n_{\sigma i} \le 1
$
is the number of $\sigma$ fermions at site $i$ (number filling) in such a way that 
$
N_\sigma = \sum_i n_{\sigma i}
$
gives the total number of $\sigma$ fermions. 
Equations~(\ref{eqn:bdg.matrix}-~\ref{eqn:number}) correspond to the 
generalization of the BdG equations~\cite{iskinbdg} to the case of spin-orbit 
coupled Fermi gases on optical lattices.

\subsection{Circulating mass currents}
\label{sec:currents}

Once we obtain the self-consistent solutions for the quasiparticle wave functions 
and the corresponding energy spectrum, it is straightforward to calculate any of the 
desired observables. For instance, in this paper we are interested in the quantum 
mechanical probability-current (mass- or particle-current) of $\sigma$ fermions 
at site $i$ defined by 
$
\mathbf{J}_{\sigma i} = J_{\sigma i}^x \hat{x}  + J_{\sigma i}^y \hat{y},
$
where
$
J_{\sigma i}^e = - i t_{\sigma\sigma}^{i,i+\hat{e}} \langle c_{\sigma i}^\dagger c_{\sigma, i+\hat{e}} - H.c. \rangle
$
is the $e$th component. Using the Bogoliubov-Valatin transformation, we obtain
\begin{align}
\label{eqn:current}
J_{\sigma i}^e = -i t_{\sigma\sigma}^{i,i+\hat{e}} \sum_n  \Big[ & (u_{n i}^\sigma)^* u_{n, i+\hat{e}}^\sigma f(\varepsilon_n) \nonumber \\
&+ (v_{n i}^\sigma)^* v_{n, i + \hat{e}}^\sigma f(-\varepsilon_n)  - H.c. \Big],
\end{align}
where we set the lattice spacing and $\hbar$ to unity.
In our two-dimensional system, we expect mass currents to circulate around the central site, 
so that
$
\mathbf{J}_{\sigma i} = J_{\sigma i} \hat{\theta},
$
where $\theta$ is the azimuthal angle. In addition, the local angular-momentum associated 
with such a circulating mass current is in the $z$ direction, and its magnitude at a particular 
site $i$ is simply related to the local mass current by 
$
L_{\sigma i}^z = x_i  J_{\sigma i}^y - y_i J_{\sigma i}^x,
$
where $x_i$ and $y_i$ are the coordinates of site $i$ with 
respect to the center, i.e. minimum of the trapping potential. 
In this paper, we are interested in the total absolute angular momentum per $\sigma$ 
particle which is given by 
$
L_\sigma^z = (1/N_\sigma) \sum_i r_i |J_{\sigma i}|.
$

Having established the BdG formalism, now we are ready to present our numerical 
solutions for the ground-state and finite-temperature phases, which are obtained 
by solving Eqs.~(\ref{eqn:bdg.matrix}-~\ref{eqn:number}) in a self-consistent manner.

\section{Numerical results}
\label{sec:numerics}
Our numerical calculations are performed on a two-dimensional 
square lattice with $L = 48$ lattice sites in each direction. 
We take $V_0 \approx 0.014 t$ as the strength of the trapping potential, and discuss both 
symmetric ($\alpha = \beta$) and asymmetric ($\alpha \ne \beta$) SOC fields. 
Since we are mainly interested in the low-filling population-balanced systems, 
we set $N_\uparrow = N_\downarrow = 125$, and study the resultant phases 
as a function of $g$, $\alpha$ and $\beta$. However, we briefly comment on the 
effects of high filling and population imbalance on the ground-state phases 
toward the end of this section. 

\begin{figure}[htb]
\centerline{\scalebox{0.58}{\includegraphics{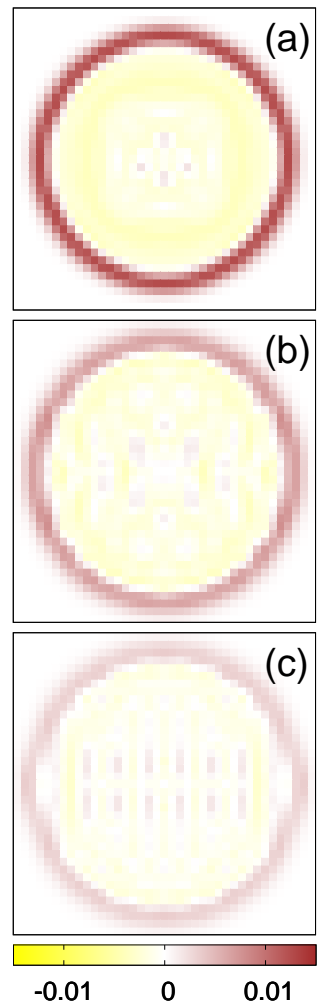}}}
\caption{\label{fig:current1} (Color online)
The strength of the $\uparrow$ mass current $\mathbf{J_{\uparrow i}}$ (in units of $t$) 
is shown as a function of lattice coordinates $x_i$ and $y_i$ at zero temperature ($T=0$) for 
noninteracting systems ($g = 0$). Here, we consider (a) a symmetric SOC
where we set $\alpha = \beta = \pi/4$, and (b-c) asymmetric SOCs where we set 
$\beta$ to $\pi/8$ in (b) and $\pi/16$ in (c), while keeping $\alpha = \pi/4$.
The lighter yellow (darker red) indicates clockwise (counterclockwise) circulation. 
$\mathbf{J_{\downarrow i}}$ has exactly the opposite circulation and it is not shown.
}
\end{figure}

In Fig.~\ref{fig:current1}, we show the strength of the $\uparrow$ mass current  
$\mathbf{J_{\uparrow i}}$ at zero temperature ($T = 0$) for noninteracting ($g = 0$) 
systems with symmetric and asymmetric SOC, where we set $\alpha = \pi/4$ 
and vary $\beta$ parameter. 
We use Fig.~\ref{fig:current1}(a) as our reference data in all of our discussions below.
First of all, the coarse-grained data near the trap center is a finite-size lattice effect 
and it does not affect our main findings. In addition, due to the time-reversal 
symmetry, $\mathbf{J_{\downarrow i}}$ has exactly the opposite circulation for 
population-balanced systems and this current is not shown throughout this paper.
As $\beta$ decreases from the symmetric case with $\beta = \pi/4$ to $\beta = 0$,
the asymmetry between $x$ and $y$ directions increase in such a way that the 
spin-flipping hopping gradually decreases to zero in the $y$ direction, and
the SOC field eventually becomes purely one dimensional in the $x$ direction. 
We find that the mass current flows as long as the SOC field has both $x$ 
and $y$ components, and its strength decreases gradually as a function of 
increasing the asymmetry of SOC fields. Note that since the asymmetric 
SOC breaks the $C_4$ symmetry of the square lattice, the resultant mass 
currents also have reduced symmetry in Figs.~\ref{fig:current1}(b) and (c). 
In addition, we see that the gas expands a little bit with increasing asymmetry, 
which is due to the decrease in density of single-particle states.

\begin{figure}[htb]
\centerline{\scalebox{0.45}{\includegraphics{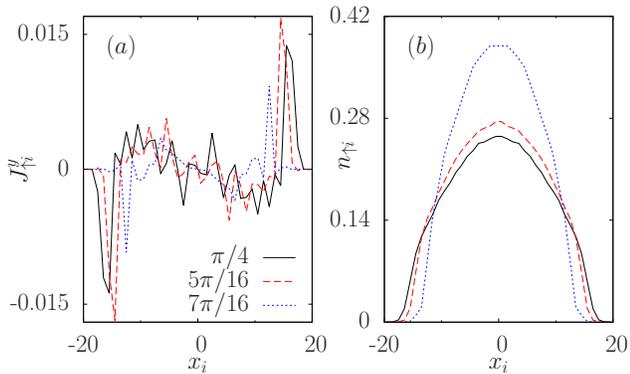}}}
\caption{\label{fig:current2} (Color online)
(a) The strength of the $\uparrow$ mass current $J_{\uparrow i}^y$ (in units of $t$),
and (b) the filling of $\uparrow$ fermions $n_{\uparrow i}$ are shown 
at $T = 0$ as a function of $x_i$, when $y_i = 0$. Here, the system is noninteracting, 
and we set $\alpha = \beta$ to $\pi/4$ (solid black), $5\pi/16$ (dashed red) 
and $7\pi/16$ (dotted blue). Coarse-grained data is a finite-size lattice effect.
}
\end{figure}

In Fig.~\ref{fig:current2}, we show the mass current $J_{\uparrow i}^y$ and filling 
$n_{\uparrow i}$ of $\uparrow$ fermions at $T = 0$ for noninteracting systems 
with varying symmetric SOC strengths $\alpha = \beta$. 
We find that $J_{\uparrow i}$ first increases 
and then decreases as $\alpha = \beta$ is increased from $0$ to $\pi/2$. 
It is expected that the strength of the mass current rapidly increases as a function of 
increasing spin-flipping hopping with respect to the spin-conserving ones, 
since the presence of a nonzero SOC is what allowed the mass current to form
in the first place. However, as the SOC terms dominate, the chemical potential
gradually drops below the band minimum, and then the system gradually crosses over
to the molecular BEC side. We note that in contrast to the no-SOC case
where the formation of a two-body bound state between $\uparrow$ and $\downarrow$ 
fermions requires a finite interaction threshold in a two-dimensional lattice, 
here it can form even for arbitrarily small interactions simply by increasing 
the SOC, due to the increased single-particle density of states.
Therefore, it is expected that the rapid increase in the strength of the mass 
current is followed by a gradual decrease at larger SOC, and that the mass
current eventually vanishes at sufficiently large SOC.
In addition, we see that the gas shrinks with increasing SOC, which is 
again due to the increased density of states. Note that the ratio between the 
trapping potential and the effective hopping terms also increases 
as $\alpha = \beta$ increases from 0 to $\pi/2$.
We also calculate the absolute angular momentum per particle $L_\sigma^z$ 
and find approximately $0.53$, $0.56$ and $0.38$ (in units of $\hbar$), when
$\alpha = \beta$ is set to $\pi/4$, $5\pi/16$ and $7\pi/16$, respectively.

\begin{figure}[htb]
\centerline{\scalebox{0.45}{\includegraphics{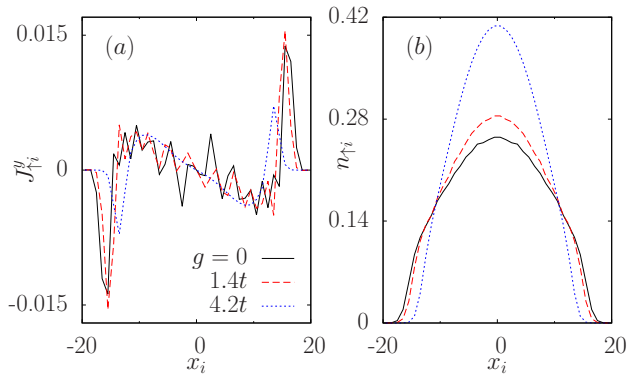}}}
\caption{\label{fig:current3} (Color online)
(a) The strength of the $\uparrow$ mass current $J_{\uparrow i}$ (in units of $t$),
and (b) the filling of $\uparrow$ fermions $n_{\uparrow i}$ are shown 
at $T = 0$ as a function of $x_i$, when $y_i = 0$. Here, the SOC is symmetric with 
$\alpha = \beta = \pi/4$, and we set $g$ to $0$ (solid black) $1.4t$ (dashed red) 
and $4.2t$ (dotted blue). Coarse-grained data is a finite-size lattice effect.
}
\end{figure}

In Fig.~\ref{fig:current3}, we show the mass current $J_{\uparrow i}^y$ and filling 
$n_{\uparrow i}$ of $\uparrow$ fermions at $T = 0$ for symmetric ($\alpha = \beta = \pi/4$) SOC 
with varying interaction strength $g$. The corresponding superfluid order parameter
$|\Delta_i|$ has an approximately peak value of $0$, $0.05t$ and $1.4t$ at 
the trap center when $g$ equals to $0$, $1.4 t$ and $4.2t$, respectively (not shown).
It is clearly seen that increasing the $g$ has qualitatively the same effect on the 
ground-state phases as increasing the SOC. For instance,
we find that the mass current first increases and then decreases as a function of
increasing $g$. In addition, the gas shrinks with increasing $g$, due to the 
formation of more tightly-bound Cooper pairs. 
We again calculate the absolute angular momentum per particle $L_\sigma^z$ 
and find approximately $0.53$, $0.49$ and $0.22$ (in units of $\hbar$), when
$g$ is set to $0$, $1.4t$ and $4.2 t$, respectively.
Therefore, $L_\sigma^z$ again deviates significantly from $\approx 0.5$ as 
a result of increased filling around the trap center. 

\begin{figure}[htb]
\centerline{\scalebox{0.45}{\includegraphics{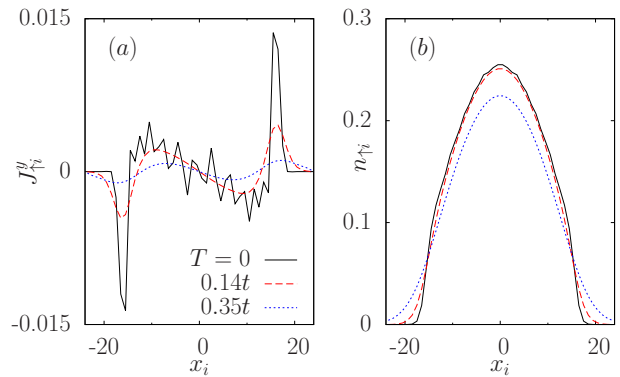}}}
\caption{\label{fig:current4} (Color online)
(a) The strength of the $\uparrow$ mass current $J_{\uparrow i}$ (in units of $t$),
and (b) the filling of $\uparrow$ fermions $n_{\uparrow i}$ are shown 
at finite $T$ as a function of $x_i$, when $y_i = 0$. Here, the system is 
noninteracting and SOC is symmetric with $\alpha = \beta = \pi/4$, and 
we set $T$ to $0$ (solid black) $0.14t$ (dashed red) 
and $0.35t$ (dotted blue). Coarse-grained data is a finite-size lattice effect.
}
\end{figure}

In Fig.~\ref{fig:current4}, we show the mass current $J_{\uparrow i}^y$ and filling 
$n_{\uparrow i}$ of $\uparrow$ fermions for noninteracting systems with 
varying temperature $T$, where we set symmetric ($\alpha = \beta = \pi/4$) SOC. 
Since finite temperature excites more and more particles to higher oscillator states 
as a function of increasing $T$, and this naturally leads to an increase in the 
system size, and decrease in the strength of the mass currents. This is clearly
seen in the figure, where the peak value of $J_{\uparrow i}^y$ reduces 
approximately to half of its $T = 0$ value when $T \simeq 0.14t$, but with a larger width. 
We calculate the absolute angular momentum per particle $L_\sigma^z$ 
and find approximately $0.53$, $0.33$ and $0.16$ (in units of $\hbar$), when
$T$ is set to $0$, $0.14t$ and $0.35t$, respectively.
Therefore, $L_\sigma^z$ decreases significantly from its ground-state value 
$\approx 0.5$ as the temperature increases. 

So far we only discussed population-balanced systems with a low filling at the
trap center, i.e. $n_{\sigma i} \lesssim 0.4$, and here we briefly comment on 
the effects of high filling and population imbalance on the ground-state phases~\cite{iskinbdg}.
First of all, due to the particle-hole symmetry of the Hamiltonian around 
$n_{\sigma i} = 0.5$, in addition to the mass-current peak circulating in the vicinity of the 
trap edge, an additional peak that is circulating mostly near the trap center 
is induced when the trap center is close to a band insulator, i.e. $n_{\sigma i} \simeq 1$. 
Therefore, mass currents show a double-ring structure in high-filling lattice systems.
Second, the strengths of mass currents are quite robust against the effects of 
imbalanced populations. For instance, when the total number of 
$\downarrow$ fermions is reduced from $N_\downarrow = 125$ in Fig.~\ref{fig:current1}(a)
to $N_\downarrow = 20$ while keeping $N_\uparrow = 125$ fixed, 
so that $n_{\uparrow} \approx 0.3$ and $n_{\downarrow} \approx 0.05$ 
at the trap center, the maximum currents 
$J_{\uparrow} \gtrsim J_{\downarrow} \approx 5 \times 10^{-3} t$ 
occur about the same radial distance as the population-balanced 
case shown in Fig.~\ref{fig:current1}(a).

\section{Discussion}
\label{sec:disc}

Having established the qualitative behavior of spontaneously induced counterflowing 
mass currents in spin-orbit coupled Fermi gases, next we argue that the origin of 
these currents can be understood via a direct correspondence with the chiral $p$-wave 
superfluids/superconductors.

\subsection{Correspondence with chiral $p$-wave systems}
\label{sec:pwave}

The origin of spontaneously induced counterflowing mass currents near the 
trap edge can be understood via a direct correspondence with the 
$p_x + ip_y$-superfluids/superconductors, e.g. $A$ phase of liquid $^3$He, 
for which it is known that a spontaneous mass current also flows near the 
boundary~\cite{stone, mizushima}. In these $p$-wave systems, the mass current 
and hence the macroscopic angular momentum is associated with the chirality 
of Cooper pairs. The chirality can be most easily seen by noting that the 
chiral $p$-wave order parameter 
$
\Delta_\mathbf{k} \propto (\hat{x} \pm i\hat{y}) \cdot \mathbf{k},
$
where $\mathbf{k}$ is the relative momentum of a Cooper pair, is an 
eigenfunction of the orbital angular momentum with eigenvalue $\pm \hbar$.
This mechanism explains our findings since it can be shown that the
order parameter of spin-orbit coupled Fermi gases with Rashba-type
SOC and $s$-wave contact interactions has chiral $p$-wave symmetry. 
To see this one needs to transform the Hamiltonian to the 
helicity basis, i.e. the pseudo-spin is no longer a good quantum number
in the presence of a SOC, and the spin direction is either parallel or 
anti-parallel to the direction of in-plane momentum in the $\pm$ 
helicity basis. In the helicity-basis representation, it becomes clear that 
only intraband Cooper pairing occurs between fermions with same helicity, 
and that the order parameter for $\pm$ helicity pairing has $p_x \mp ip_y$ 
symmetry~\cite{zhai}.

Although the basic mechanism in chiral $p$-wave systems and spin-orbit 
coupled Fermi gases share some similarities with respect to spontaneously 
induced edge currents, they also differ in fundamental ways. For instance, 
in contrast to the chiral $p$-wave systems where the formation of Cooper 
pairs and hence the mass current requires an interacting system, 
in spin-orbit coupled Fermi gases currents flow even in the absence of interactions. 
In addition, the chiral $p$-wave systems belong to the topological class of 
integer quantum Hall systems, and since these systems both break time-reversal 
symmetry, they exhibit spontaneous edge currents circulating along a particular 
direction. However, spin-orbit coupled Fermi gases preserve time-reversal 
symmetry just like quantum spin Hall systems, and therefore, they both exhibit 
spontaneously induced counterflowing $\uparrow$ and $\downarrow$ edge
currents. Next, we comment on the experimental realization of spontaneously 
induced edge currents in atomic systems.

\subsection{Experimental realization}
\label{sec:exp}

In comparison to bulk (center) effects, there is no doubt that the edge effects are 
more difficult to observe in trapped atomic systems. This is because as the local
density decreases from the central region to the edges, it gets harder to probe
local properties. In addition, the local critical superfluid temperature eventually 
drops below the temperature of the system near the edges, and the superfluidity 
is also lost starting from the edges inwards. One way to partially overcome such 
problems could be to load the optical lattice potential with higher particle fillings such 
that the trap center is close to a band insulator, i.e. unit filling. 
In this case, an additional mass-current peak circulates mostly near the trap 
center (see above), and such a local current could be easier to probe 
with current experimental capabilities.

The measurement of the angular momentum of rotating 
atomic systems have so far only been achieved indirectly, by observing the 
shift of the radial quadrupole modes. While this technique was initially used 
for rotating atomic BECs~\cite{chevy, cornell}, it has recently been applied 
to the rotating fermionic superfluids in the BCS-BEC crossover~\cite{grimm}.
We believe a similar technique could be used for measuring the intrinsic 
angular momentum of spin-orbit coupled Fermi gases, which may provide 
an indirect evidence for counterflowing mass currents. In addition, we
note that a realistic scheme has recently been proposed to detect topological 
edge states in an optical lattice under a synthetic magnetic field~\cite{goldman}.
This proposal is based on a generalization of Bragg spectroscopy that is 
sensitive to angular momentum.

\section{Conclusions}
\label{sec:conc}

To conclude, we showed that the presence of a Rashba-type SOC spontaneously 
induces counterflowing $\uparrow$ and $\downarrow$ mass currents. 
While these currents have a single peak that is circulating mostly near the trap 
edge in low-filling lattice systems, an additional peak that is circulating mostly 
near the trap center is also induced in high-filling lattice systems, 
exhibiting a double-ring structure.
We note that our results for the low-filling lattice systems are directly applicable 
to the dilute systems (without the optical lattice potential), for which we expect
qualitatively similar behavior.
These currents flow even in noninteracting systems, and they are quite robust
against the effects of imbalanced populations and/or asymmetric SOC in 
$x$ and $y$ directions. However, their strength decreases toward the 
molecular BEC limit, which can be achieved either by increasing the SOC 
or the interaction strength. 
We argued that the origin of spontaneously induced counterflowing mass 
currents in spin-orbit coupled Fermi gases can be understood via a direct 
correspondence with the chiral $p$-wave superfluids/superconductors,
however, with some fundamental differences in between.

\section{Acknowledgments}
\label{sec:ack}
This work is supported by the Marie Curie IRG Grant No. FP7-PEOPLE-IRG-2010-268239, 
T\"{U}B$\dot{\mathrm{I}}$TAK Career Grant No. 3501-110T839, and 
T\"{U}BA-GEB$\dot{\mathrm{I}}$P. ED is also supported by T\"{U}B$\dot{\mathrm{I}}$TAK-2215 PhD Fellowship.

\end{document}